\RequirePackage{snapshot}
\documentclass{webofc}
\usepackage[varg]{txfonts}   
\usepackage{graphicx}
\usepackage{amsmath}
\usepackage[section]{placeins}
\usepackage{wrapfig}

\usepackage{tikz,varwidth}
\usepackage{tikzscale}
\usetikzlibrary{shapes.geometric, patterns}
\usetikzlibrary{arrows.meta,shapes,automata,petri,positioning,calc,decorations.markings}
\usepackage{tkz-euclide}

\begin{document}
\title{Emergent Structure in QCD}
%
%

\author{
  \firstname{James} \lastname{Biddle}\inst{1}\fnsep\thanks{\email{james.biddle@adelaide.edu.au}}
  \and
  \firstname{Waseem} \lastname{Kamleh}\inst{1}\fnsep\thanks{\email{waseem.kamleh@adelaide.edu.au}}
  \and
  \firstname{Derek} \lastname{Leinweber}\inst{1}\fnsep\thanks{\email{derek.leinweber@adelaide.edu.au}}
}

\institute{Centre for the Subatomic Structure of Matter, Department of Physics,\\ 
  The University of Adelaide, SA 5005, Australia}

\abstract{ The structure of the $SU(3)$ gauge-field vacuum is explored through visualisations of
  centre vortices and topological charge density. Stereoscopic visualisations highlight interesting
  features of the vortex vacuum, especially the frequency with which singular points appear and the
  important connection between branching points and topological charge. This work demonstrates how
  visualisations of the QCD ground-state fields can reveal new perspectives of centre-vortex
  structure.}
\maketitle
\section{Introduction}
\label{intro}

Quantum Chromodynamics (QCD) is the fundamental relativistic quantum
field theory underpinning the strong interactions of nature. The
gluons of QCD carry colour charge and self-couple. This self-coupling
makes the empty vacuum unstable to the formation of non-trivial quark
and gluon condensates. These non-trivial ground-state ``QCD-vacuum''
field configurations form the foundation of matter.

\begin{figure}[t]
\centering
\includegraphics[width=1.0\linewidth,clip]{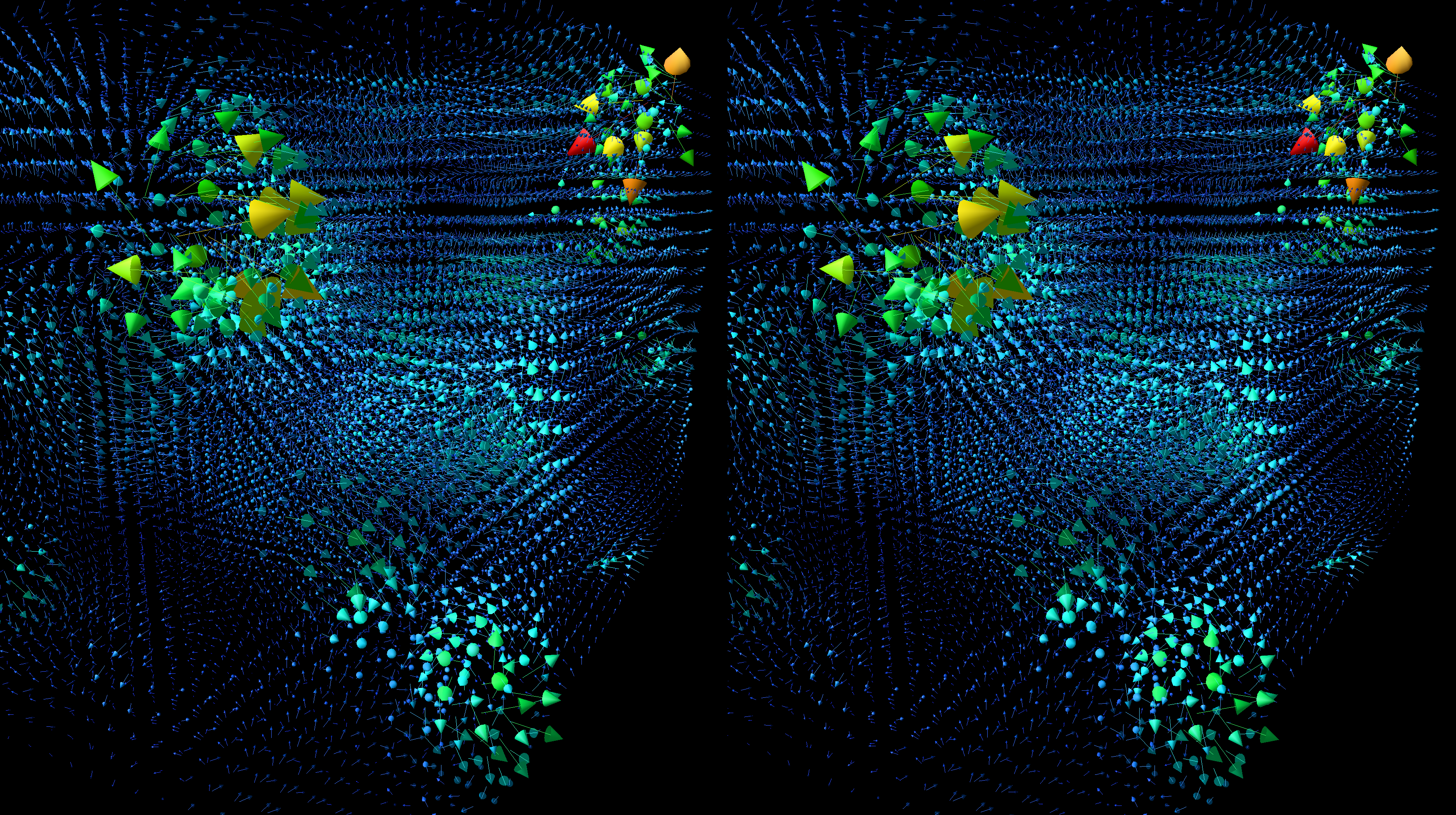}
\caption{Stereoscopic image of one of the eight chromo-magnetic fields composing the nontrivial
  vacuum of QCD.  Hints for stereoscopic viewing are provided in the text.}
\label{fig:GluonField}
\end{figure}

There are eight chromo-electric and eight chromo-magnetic fields composing the QCD vacuum. An
stereoscopic illustration of one of these chromo-magnetic fields is provided in
Fig.~\ref{fig:GluonField}.  Animations of the fields are also available
\cite{Biddle:2019abu,StructVac:2019,StructVacHD:2018}.

The essential, fundamentally-important, nonperturbative features of the QCD vacuum fields are the
dynamical generation of mass through chiral symmetry breaking, and the confinement of quarks.  But
what is the fundamental mechanism of QCD that underpins these phenomena?  What aspect of the QCD
vacuum causes quarks to be confined?  Which aspect is responsible for dynamical mass generation?
Do the underlying mechanisms share a common origin?

One of the most promising candidates is the centre vortex
perspective~\cite{'tHooft:1977hy,'tHooft:1979uj}. This perspective describes the nature
of the QCD vacuum in terms of the most fundamental centre of the $SU(3)$ gauge group, characterised
by the three values of $\sqrt[3]{1}$. Centre vortices on the lattice have been demonstrated to give
rise to indicators of confinement such as the linear static quark potential \cite{Cais:2008za,
  Langfeld:2003ev, Trewartha:2015ida,Greensite:2003bk,DelDebbio:1998luz} and enhancement of the
infrared gluon propagator~\cite{Bowman:2010zr, Biddle:2018dtc, Langfeld:2001cz,
  Quandt:2010yq}. They also reproduce indicators of dynamical chiral symmetry breaking through
enhancement of the infrared quark mass function~\cite{Bowman:2008qd, Trewartha:2015nna} and mass
splitting in the low-lying hadron spectrum~\cite{Trewartha:2017ive, Trewartha:2015nna,
  OMalley:2011aa}.

As such, it is interesting to ask, what do these structures look like? To this end, we present
visualisations of centre vortices and topological charge density as identified on lattice
gauge-field configurations. These visualisations are presented as stereoscopic images.
To see the 3D image, try the following:
\begin{enumerate}
\item If you are viewing the image on a monitor, ensure the image width is 12 to 13 cm.
\item Bring your eyes very close to one of the image pairs.
\item Close your eyes and relax.
\item Open your eyes and allow the (blurry) images to line up. Tilting your head from side to side
  will move the images vertically. 
\item Move back slowly until your eyes are able to focus.  There's no need to cross your eyes!
\end{enumerate}

\begin{wrapfigure}{r}{0.4\linewidth}
  \centering
  \includegraphics[width=3.0cm]{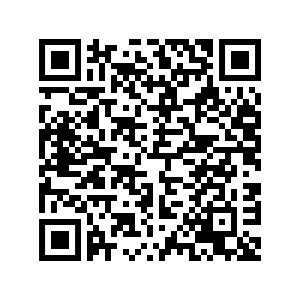}\\
  \vspace{-10pt}
  \caption{\label{fig:QRCodes}QR code to download the augmented reality app.}
\end{wrapfigure}

This work also features Josh Charvetto's augmented reality app for lattice QCD. Get the app using
the QR code in Fig.~\ref{fig:QRCodes} and see a whole new view of these visualisations through the
camera's eye.

\section{Finding Centre Vortices}
\label{sec-1}

Centre vortices are identified through a gauge fixing procedure designed to bring the lattice link
variables as close as possible to the identity matrix multiplied by a phase equal to one of the
three cube-roots of one \cite{DelDebbio:1996mh,Langfeld:1997jx,Langfeld:2003ev}.
\clearpage
\noindent
The links $U_\mu(x)$ are then projected to the centre elements of $SU(3)$,
\begin{equation}
U_\mu(x) \to Z_\mu(x)\, , \quad \textrm{where }
Z_\mu(x) = e^{ 2 \pi i\, {m_\mu(x)}/{3} } \, \mathbf{I} \, ,
\quad\textrm{and } m_\mu(x) = -1, 0, 1\, ,
\end{equation}
such that the gluon field is characterised by the most fundamental aspect of the $SU(3)$ link
variable, the centre.  
This ``vortex-only'' field, $Z_\mu(x)$, can be examined to learn the extent to which centre
vortices alone capture the essence of nonperturbative QCD.

Vortices are identified by the centre charge, $z$, given
by the product of the vortex-only field around an elementary square or plaquette on the lattice
\begin{equation}
z = \frac{1}{3}\, \mbox{Tr}\, \prod_\Box Z_\mu(x) = e^{ 2 \pi i\, {m}/{3} } \, .
\end{equation}
%
If $z \neq 1$, a vortex with charge $z$ characterised by $m$ pierces the plaquette.
%

\section{Centre Vortex Features}

\begin{figure}[t]
  \centering
  \includegraphics[width=1.0\linewidth]{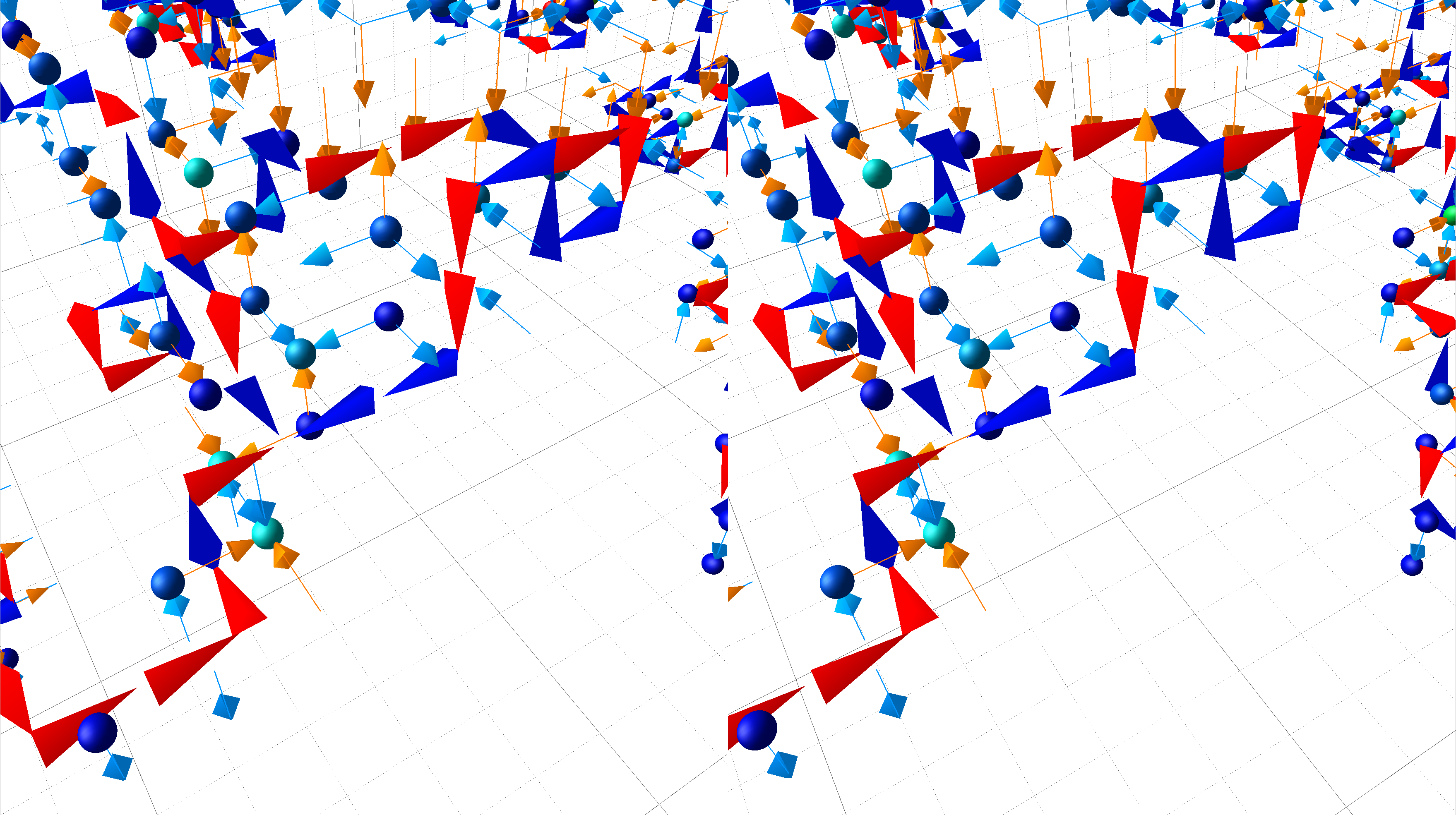}
  \caption{Stereoscopic image of centre vortices as identified on the
    lattice. Vortex features including vortex lines (jets), branching
    points (3-jet combinations), crossing points (4 jets), indicator
    links (arrows) and singular points (spheres) are described in the
    text.}
  \label{fig:Vortices}
\end{figure}

{\bf Vortex Lines:} The plaquettes with nontrivial centre charge are
plotted as red or blue jets 
\begin{wrapfigure}{r}{0.4\linewidth}
  \centering
  \vspace{-6pt}
  \includegraphics[width=\linewidth]{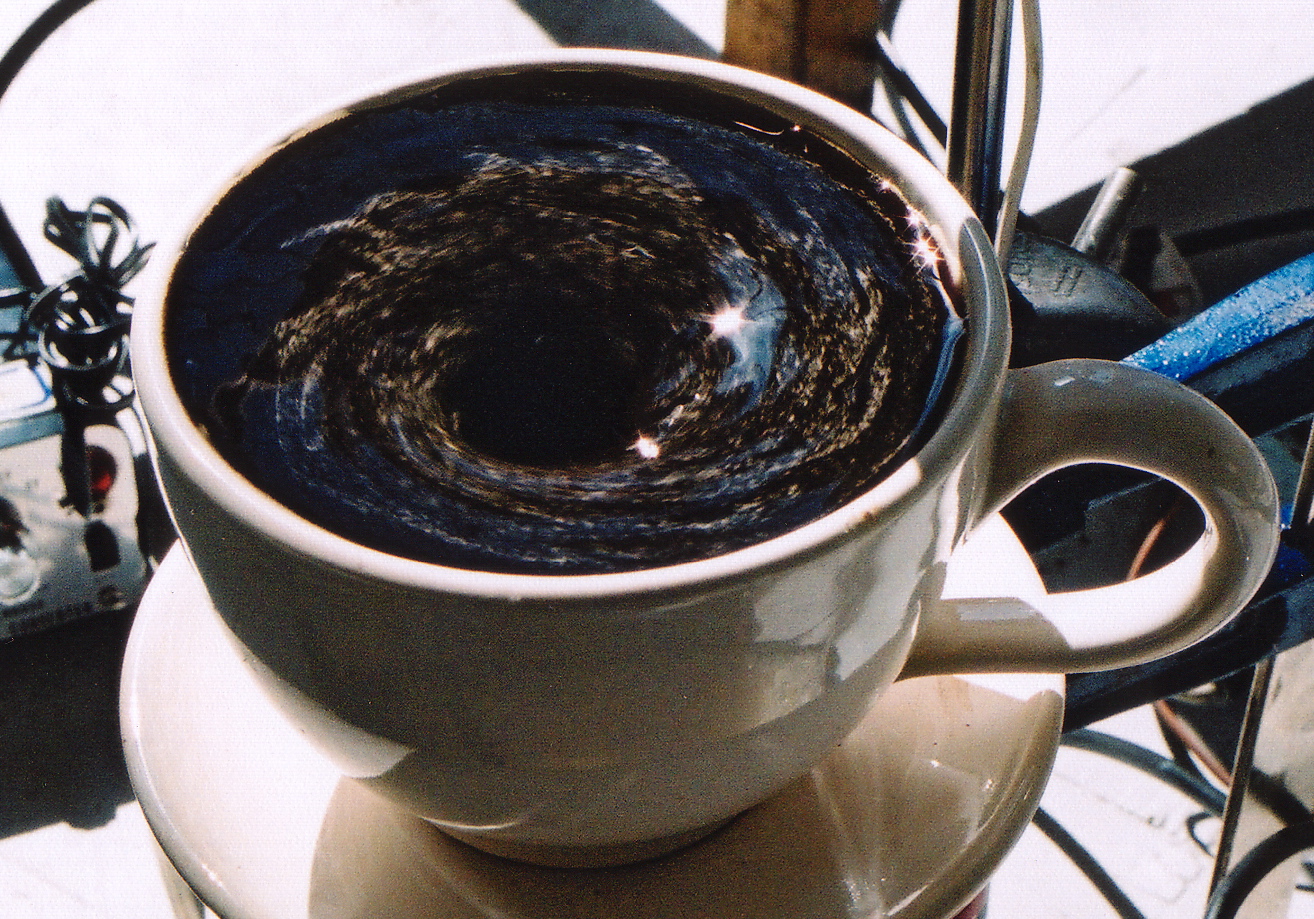}
  \vspace{-24pt}
  \caption{\label{fig:VortexExample} An example of a familiar vortex.
    The position of the vortex is characterised by the line running
    along the centre of the rotation.}
\end{wrapfigure}
piercing the centre of the plaquette in Figure~\ref{fig:Vortices}.  Here we are considering a 3D
slice of the 4D space-time lattice at fixed time.  The orientation and colour of the jets reflect
the value of the non-trivial centre charge.  Using a right-hand rule for the direction, plaquettes
with $m=+1$ are illustrated by blue jets in the forward direction, and plaquettes with $m=-1$ are
illustrated by red jets in the backward direction.  Thus, the jets show the directed flow of centre
charge $z=e^{2\pi i/3}$ through spatial plaquettes.  They are analogous to the line running down
the centre of a vortex as illustrated in Fig.~\ref{fig:VortexExample}.

Figure~\ref{fig:Vortices} exhibits rich emergent structure in the
nonperturbative QCD ground-state fields.
\vspace{12pt}

{\bf Branching Points or Monopoles:} In $SU(3)$ gauge theory, three vortex lines can merge
into or emerge from a single point.  Their prevalence is surprising, as is their correlation with topological charge density.

\bigskip {\bf Vortex Sheet Indicator Links:} As the vortex line moves in time, it creates
a vortex sheet in 4D spacetime.  This movement is illustrated by arrows along the links of
the lattice (shown as cyan and orange arrows in Fig.~\ref{fig:Vortices} indicating centre
charge flowing through the suppressed time direction.

\bigskip {\bf Singular Points:} When the vortex sheet spans all four space-time
dimensions, it can generate topological charge. Lattice sites with this property are called
singular points and are illustrated by spheres. The sphere colour indicates the number of
times the sheet adjacent to a point can generate a topological charge
contribution \cite{Biddle:2019gke}.

\begin{figure}
  \centering
  \includegraphics[width=1.0\linewidth]{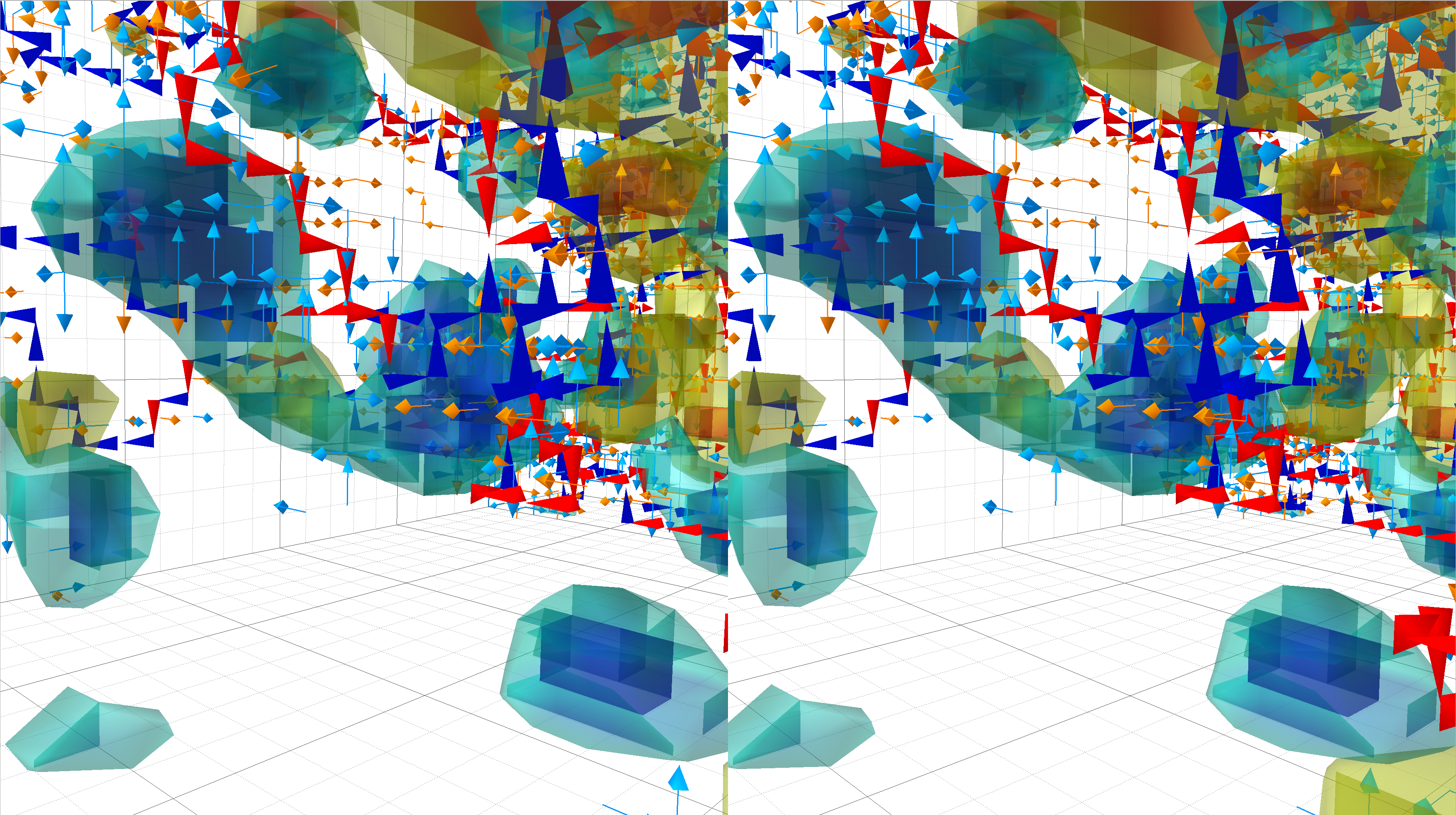}
  \caption{Stereoscopic image illustrating the correlation of
    topological charge density in the gluon field with the positions
    of centre vortices.  Positive charge density is rendered red
    through yellow and negative charge density is rendered blue
    through blue.}
  \label{fig:TopQ}
\end{figure}

\bigskip {\bf Topological Charge Density:} Non-trivial topological charge is often associated with
instanton-like field configurations which dynamically generate the mass the proton and other
hadrons. Topological charge density is plotted as red through yellow volumes for positive charge
density and blue through green volumes for negative charge density in Fig.~\ref{fig:TopQ}.

\bigskip {\bf Vortices \& Topological Charge:} Vortices are somewhat correlated with the
positions of significant topological charge density, but not in a strong manner.  However,
the percolation of vortex structure is significant and the removal of these vortices
destroys most instanton-like objects.

\section{Conclusion}

Visualisations of centre-vortex phenomena provide a fascinating glimpse into the structure of the
QCD ground-state fields.  Patterns and correlations that remain hidden within giga-bytes of data
become clear as one examines the data drawing on our highly-evolved sense of vision.  It provides
an extremely powerful form of data analysis.  Animations of centre-vortex structure are also
available \cite{Biddle:2019abu,Vortices:2019,VorticesHD:2018}.

A more detailed presentation of this work providing access to the full 3D models is available in
Ref.~\cite{Biddle:2019gke}.  It is interesting to reflect on the discoveries made in
Ref.~\cite{Biddle:2019gke} through visualisation. These include:

\begin{itemize}
\item The high density of centre vortices and the complexity of their structure.

\item The long-distance structure of the centre vortices as they percolate through the vacuum fields.

\item The proliferation of branching points in $SU(3)$ gauge theory.

\item The correlation of branching points with topological-charge-density peaks, revealing the vital
      role of vortex branching in generating topological charge density.

\item Understanding the sheet-like structure of centre vortices in four dimensions as they pierce one 3D
      time-slice of the lattice and traverse to the next.

\item The speed with which centre vortices move through the lattice.

\item The rapid loss of centre vortices under gauge-field smoothing routines.

\end{itemize}
The new understanding of centre-vortex branching and topological charge is important.  This
mechanism governs the size of instanton-like field configurations which generate a density of near
zero-modes in the Dirac operator, dynamically breaking chiral symmetry in the ground-state fields
and generating the mass of visible matter.

\section*{Acknowledgements}

We thank Daniel Trewartha for his contributions to the gauge ensembles
underlying this investigation.  This research is supported with
supercomputing resources provided by the Phoenix HPC service at the
University of Adelaide and the National Computational Infrastructure
(NCI) supported by the Australian Government. This research is
supported by the Australian Research Council through Grants
No. DP190102215, DP150103164, DP190100297 and LE190100021.

\providecommand{\href}[2]{#2}\begingroup\raggedright\endgroup

\end{document}